\documentclass[aps,prl,twocolumn,superscriptaddress,showpacs,amsmath,amssymb,nofootinbib]{revtex4}
\usepackage{graphicx,epsf,amsmath}
\usepackage[]{latexsym}
\newcommand{\be}{\begin{eqnarray}}
\newcommand{\ee}{\end{eqnarray}}

\newcommand{\ket}[1]{\mbox{$\mid #1\,\rangle$}}

\newcommand{\pro}[2]{\mbox{$\langle\, #1 \mid #2\,\rangle$}}

\renewcommand{\d}{\mbox{${\rm d}$}}
\newcommand{\lp}{\ell_{\rm p}}
\newcommand{\mpl}{m_{\rm p}}
\newcommand{\gn}{G_{\rm N}}
\newcommand{\rh}{R_{\rm H}}
\begin{document}
%
%
\title{Quantum hoop conjecture: Black hole formation by particle collisions}
\author{Roberto~Casadio}
\email{casadio@bo.infn.it}
\affiliation{Dipartimento di Fisica e Astronomia, Universit\`a di Bologna,
via~Irnerio~46, 40126~Bologna, Italy}
\affiliation{I.N.F.N., Sezione di Bologna, viale Berti~Pichat~6/2, 40127~Bologna, Italy}
\author{Octavian~Micu}
\email{octavian.micu@spacescience.ro}
\affiliation{Institute of Space Science, Bucharest,
P.O.~Box MG-23, RO-077125 Bucharest-Magurele, Romania}
\author{Fabio Scardigli}
\email{fabio@phys.ntu.edu.tw}
\affiliation{Dipartimento di Matematica, Politecnico di Milano, Piazza L.~da~Vinci~32,
20133~Milano, Italy}
\affiliation{Yukawa Institute for Theoretical Physics,
Kyoto University, Kyoto 606-8502, Japan}
\begin{abstract}
We address the issue of (quantum) black hole formation by particle collision in quantum
physics.
We start by constructing the horizon wave-function for quantum mechanical
states representing two highly boosted non-interacting particles that collide in flat
one-dimensional space.
From this wave-function, we then derive a probability that the system becomes a black hole
as a function of the initial momenta and spatial separation between the particles.
This probability allows us to extend the hoop conjecture to quantum mechanics and
estimate corrections to its classical counterpart.
\end{abstract}
\maketitle
\section{Introduction}
The general relativistic (GR) description of the gravitational collapse, leading to the formation
of black holes (BHs), was first investigated in the seminal papers of Oppenheimer and
co-workers~\cite{OS}, but a thorough understanding of the physics of such processes still
stands as one of the most challenging issues for contemporary theoretical physics.
The literature on the subject has grown immensely (see, e.g.~Ref.~\cite{joshi}),
but many technical and conceptual difficulties remain unsolved, particularly if one tries
to account for the quantum mechanical (QM) nature of collapsing matter.
What is unanimously accepted is that the gravitational interaction becomes important whenever
a large enough amount of matter is ``compacted'' within a sufficiently small volume.
K.~Thorne formulated this idea in the {\em hoop conjecture\/}~\cite{Thorne:1972ji},
which states that a BH forms if two colliding objects fall within their ``black disk''.
Assuming the final configuration is (approximately) spherically symmetric, this occurs when
the system occupies a sphere whose radius $r$ is smaller than the gravitational
Schwarzschild radius,
\be
r
\lesssim
\rh
\equiv
2\,\lp\,\frac{E}{\mpl}
\ ,
\label{hoop}
\ee
where $E$ is the total energy in the centre-of-mass frame (see next Section for more details).
Note that we use units with $c=1$, the Newton constant $\gn=\lp/\mpl$, where $\lp$ and $\mpl$
are the Planck length and mass, respectively, and $\hbar=\lp\,\mpl$~\footnote{These
units make it apparent that $\gn$ converts mass into length, thus providing
a natural link between energy and positions.}.
\par
The hoop conjecture applies to astrophysical bodies, whose energy is orders
of magnitude above the scale of quantum gravity, and can therefore be reasonably described
by classical GR~\cite{joshi,Thorne:1972ji,payne,murchadha}.
One of the most important questions which then arises is what happens 
when the total energy of the colliding particles is of the Planck size or {\em less\/}~\cite{qgc}.
Just to give this question a precise meaning is a conceptual challenge, because QM effects
may hardly be neglected~\cite{acmo}, and the very notion of horizon becomes ``fuzzy''. 
In fact, it was recently proposed in Refs.~\cite{Casadio} to define
a wave-function for the horizon, which can be associated with any localised QM particle.
The auxiliary wave-function yields the probability of finding a horizon of a certain radius centred
around the source, and one can therefore determine the probability that a QM
particle is a BH depending on its mass.
This probability is found to vanish very fast for particles lighter than the Planck
mass, as one expects from qualitative arguments.
\par
We remark that a realistic description of quantum (with $E\simeq\mpl$)~\cite{hsu} or classical 
($E\gg \mpl$) BHs very likely requires the knowledge of their microscopic structure~\cite{dvali}.
We however do not consider such important details here, and just address the conceptual
problem of developing a framework which can be used to study the formation of horizons 
in systems containing QM~sources.
Of course, a more canonical framework already exists, in principle,
and is given by quantum field theory on curved backgrounds coupled to the semiclassical
Einstein equations~\cite{pina}.
Thereby, one should be able to describe quantum matter states on a sufficiently arbitrary
space-time, which is to be determined self-consistently by solving the Einstein equations
with the corresponding renormalised matter energy-momentum tensor.
Since obtaining the normal modes and building the matter Fock space is
in general impossible, this procedure has failed to provide practical estimates
so far~\footnote{Computing the back-reaction of Hawking radiation on a BH
space-time is the typical example of such failures.}.
\par
In this work, after reviewing the case of a single spherically symmetric particle,
we shall consider two-particle QM states and build their horizon wave-function.
This construction will naturally lead to a QM generalisation of the hoop conjecture
and specific corrections to its classical formulation~\eqref{hoop}.
It is important to remark from the onset that these results will be obtained
analytically, but at the price of making several rather strong simplifying assumptions.
In particular, we shall just consider free particles in one spatial dimension, and neglect
any space-time curvature. 
\section{Horizon wave-function in spherical symmetry}
Inspired by Eq.~\eqref{hoop}, we can define a horizon wave-function given the QM
wave-function of a particle in position space~\cite{Casadio}.
The idea stems from the classical GR theory of spherically symmetric systems,
for which the metric $g_{\mu\nu}$ can always be written as
\be
\d s^2
=
g_{ij}\,\d x^i\,\d x^j
+
r^2(x^i)\left(\d\theta^2+\sin^2\theta\,\d\phi^2\right)
\ ,
\label{metric}
\ee
with $x^i=(x^1,x^2)$ coordinates on surfaces where the angles $\theta$ and $\phi$ are constant.
The location of a trapping horizon, a surface where the escape velocity equals
the speed of light, is determined by the equation~\cite{murchadha}
\be
0
=
g^{ij}\,\nabla_i r\,\nabla_j r
=
1-\frac{2\,M}{r}
\ ,
\label{th}
\ee
where $\nabla_i r$ is the covector perpendicular to surfaces of constant area
$\mathcal{A}=4\,\pi\,r^2$.
The function $M=\lp\,m/\mpl$ is the active (Misner-Sharp) gravitational 
mass, representing the total energy enclosed within a sphere of radius $r$
and, if we set $x^1=t$ and $x^2=r$, we find
\be
M(t,r)=\frac{4\,\pi\,\lp}{3\,\mpl}\int_0^r \rho(t, \bar r)\,\bar r^2\,\d \bar r
\ ,
\label{M}
\ee
as if the space inside the sphere were flat.
\par
For elementary particles we know for an experimental fact that QM effects
may not be neglected~\cite{acmo}.
In fact, the Heisenberg principle of QM introduces an uncertainty
in the spatial localisation of a spinless point-like source of mass $m$, typically of
the order of the Compton-de~Broglie length,
\be
\lambda_m
\simeq
\lp\,{\mpl}/{m}
\ .
\label{lambdaM}
\ee
Assuming QM is a better description of reality implies that 
the Schwarzschild radius in Eq.~\eqref{hoop} with $E=m$ only makes sense if
$
\rh\gtrsim \lambda_m
$,
or
$
m
\gtrsim
\mpl
$
(and $M\gtrsim\lp$).
Note we employed the flat space Compton length~\eqref{lambdaM},
which is likely the particle's self-gravity will affect, but it is still a reasonable
order of magnitude estimate, and BHs can therefore only exist with mass
(much) larger than the Planck scale.
\par
Let us now consider a QM state $\psi_{\rm S}$ representing
a massive particle {\em localised in space\/} and {\em at rest\/} in the chosen
reference frame.
Having defined suitable Hamiltonian eigenmodes,
$
\hat H\,\ket{\psi_E}=E\,\ket{\psi_E}
$,
where $H$ can be specified depending on the model we wish to consider,
the state $\psi_{\rm S}$ can be decomposed as
\be
\ket{\psi_{\rm S}}
=
\sum_E\,C(E)\,\ket{\psi_E}
\ .
\label{CE}
\ee
If we further assume the particle is {\em spherically symmetric\/},
we can invert the expression of the Schwarzschild radius in Eq.~\eqref{hoop}
to obtain $E$ as a function of $\rh$.
We then define the {\em horizon wave-function\/} as
\be
\psi_{\rm H}(\rh)
\propto
C\left(\mpl\,{\rh}/{2\,\lp}\right)
\ ,
\ee
whose normalisation is finally fixed in the inner product
\be
\pro{\psi_{\rm H}}{\phi_{\rm H}}
=
4\,\pi\,\int_0^\infty
\psi_{\rm H}^*(\rh)\,\phi_{\rm H}(\rh)\,\rh^2\,\d \rh
\ .
\ee
We interpret the normalised wave-function $\psi_{\rm H}$ simply as yielding the probability
that we would detect a horizon of areal radius $r=\rh$ associated with the particle in the
QM state $\psi_{\rm S}$.
Such a horizon is necessarily ``fuzzy'', like the position of the particle itself.
The probability density that the particle lies inside its own horizon of radius $r=\rh$
will next be given by
\be
P_<(r<\rh)
=
P_{\rm S}(r<\rh)\,P_{\rm H}(\rh)
\ ,
\label{PrlessH}
\ee
where
$
P_{\rm S}(r<\rh)
=
4\,\pi\,\int_0^{\rh}
|\psi_{\rm S}(r)|^2\,r^2\,\d r
$
is the probability that the particle is inside a sphere of radius $r=\rh$,
and
$
P_{\rm H}(\rh)
=
4\,\pi\,\rh^2\,|\psi_{\rm H}(\rh)|^2
$
is the probability that the horizon is located on the sphere of radius $r=\rh$.
Finally, the probability that the particle described by the wave-function $\psi_{\rm S}$
is a BH will be obtained by integrating~\eqref{PrlessH} over all possible
values of the radius,
\be
P_{\rm BH}
=
\int_0^\infty P_<(r<\rh)\,\d \rh
\ .
\label{PBH}
\ee
The above general formulation can be easily applied to a particle described by a
spherically symmetric Gaussian wave-function, for which one obtains a vanishing 
probability that the particle is a BH when its mass is smaller than about
$\mpl/4$ (for all the details, see Refs.~\cite{Casadio}).
\section{Two-particle collisions in one dimension}
It is straightforward to extend the above construction to a state containing two
free particles in one-dimensional flat space.
We again represent each particle at the time $t=0$ and position $X_i$ ($i=1$ or $2$)
by means of Gaussian wave-functions,
\be
\pro{x_i;0}{\psi_{\rm S}^{(i)}}
\equiv
\psi_{\rm S}(x_i)
=
e^{-i\,\frac{P_i\,x_i}{\hbar}}\,\frac{e^{-\frac{\left(x_i-X_i\right)^2}{2\,\ell_i}}}{\sqrt{\pi^{1/2}\,\ell_i}}
\ ,
\ee
where $\ell_i$ is the width and $P_i$ the linear momentum
(which remain constant).
The total initial wave-function is then just the product of the two one-particle states,
\be
\pro{x_1,x_2;0}{\psi_{\rm S}^{(1,2)}}
\equiv
\psi_{\rm S}(x_1,x_2)
=
\psi_{\rm S}(x_1)\,\psi_{\rm S}(x_2)
\ .
\label{PsiProd}
\ee
Like in the one-particle case or Refs.~\cite{Casadio},
it is convenient to go through momentum space in order to compute the
spectral decomposition.
We find
\be
\pro{p_i;0}{\psi_{\rm S}^{(i)}}
\equiv
\psi_{\rm S}(p_i)
=
e^{-i\,\frac{p_i\,X_i}{\hbar}}\,
\frac{e^{-\frac{\left(p_i-P_i\right)^2}{2\,\Delta_i}}}{\sqrt{\pi^{1/2}\,\Delta_i}}
\ ,
\ee
where $\Delta_i=\hbar/\ell_i$.
For $t>0$, the components of the momentum modes for each particle
therefore evolve as
$\psi_{\rm S}(p_i;t)=e^{-i\,\frac{E_i\,t}{\hbar}}\,\psi_{\rm S}(p_i)$,
where, in the following, we shall use the flat space, relativistic dispersion relation
\be
E_i
=
\sqrt{p_i^2+m_i^2}
\ .
\label{Ep}
\ee
If the particles were at rest ($P_i=0$), we could assume $\ell_i=\lambda_{m_i}$
(and $\Delta_i=m_i$).
For realistic elementary particles $m_1\simeq m_2\ll\mpl$, and one expects
the probability of forming a BH will become significant only
for $|P_i|\sim E_i\sim \mpl$.
From
$
P_i=\frac{m_i\,v_i}{\sqrt{1-v_i^2}}
$,
we obtain
\be
\ell_i
=
\frac{\hbar}{\sqrt{P_i^2+m_i^2}}
\simeq
\frac{\lp\,\mpl}{|P_i|}
\ ,
\qquad
\Delta_i
\simeq
|P_i|
\ .
\ee
The two-particle state can now be written as
\be
\ket{\psi_{\rm S}^{(1,2)}}
=
\prod_{i=1}^2
\left[
\int\limits_{-\infty}^{+\infty}
\d p_i
\,\psi_{\rm S}(p_i,t)\,\ket{p_i}
\right]
\ ,
\label{PsiPp}
\ee
and the relevant coefficients in the spectral decomposition~\eqref{CE} are given by
the sum of all the components of the product wave-function~\eqref{PsiPp}
corresponding to the same total energy $E$.
Since we shall not be concerned with the evolution of the two-particle system here,
we can simply evaluate such coefficients at $t=0$, which yields
\be
C(E)
&\!\!=\!\!&
\int\limits_{-\infty}^{+\infty}
\int\limits_{-\infty}^{+\infty}
\psi_{\rm S}(E_1)\,\psi_{\rm S}(E_2)\,
\delta(E-E_1-E_2)\,
\d E_1\,\d E_2
\notag
\\
&\!\!=\!\!&
\int\limits_{-\infty}^{+\infty}
\int\limits_{-\infty}^{+\infty}
\psi_{\rm S}(p_1)\,\psi_{\rm S}(p_2)\,
\delta(E-E_1(p_1)-E_2(p_2))
\notag
\\
&&
\phantom{\int\int}\quad
\times
\frac{\d E_1}{\d p_1}\,\frac{\d E_2}{\d p_2}\,
\d p_1\,\d p_2
\label{C(E)}
\\
&\!\!\simeq\!\!&
\int\limits_{-\infty}^{+\infty}
\int\limits_{-\infty}^{+\infty}
\psi_{\rm S}(p_1)\,\psi_{\rm S}(p_2)\,
\delta(E-|p_1|-|p_2|)
\,
\d p_1\,\d p_2
\ ,
\notag
\ee
where we used $\d E_i\simeq \d p_i$ for $|p_i|\sim |P_i|\gg m_i$.
\par
The horizon wave-function must be computed
in the centre-of-mass frame of the two-particle system,
so that
\be
P_1=-P_2
\equiv P>0
\ .
\ee
From $P\sim\mpl\gg m_1\simeq m_2$, we can also set
\be
X_1\simeq -X_2
\equiv 
X>0
\ .
\label{Xcm}
\ee
After replacing the expression of the Schwarzschild radius from Eq.~\eqref{hoop}
into Eq.~\eqref{C(E)}, we obtain the unnormalised wave-function
\be
\psi_{\rm H}(\rh)
&\!\!\propto\!\!&
e^{-\frac{\mpl\rh^2}{16\lp^2\,P}-\frac{X^2P^2}{\lp^2\mpl^2}}
{\rm Erf}\left(1+\frac{\mpl\,\rh}{4\,\lp\,P}+i\,\frac{XP}{\lp\,\mpl}\right)
\nonumber
\\
&&
-e^{-\frac{\mpl\rh^2}{16\lp^2\,P}-\frac{X^2P^2}{\lp^2\mpl^2}}
{\rm Erf}\left(1-\frac{\mpl\,\rh}{4\,\lp\,P}-i\,\frac{XP}{\lp\,\mpl}\right)
\nonumber
\\
&&
+
2\,e^{-1-\frac{2 i X P}{\lp \mpl}-\frac{\mpl \rh^2}{16 \lp^2 P}}
\cosh\left(\frac{\mpl\rh}{2\,\lp P}
+i\frac{\rh X}{2\,\lp^2}\right)
\nonumber
\\
&&
\times\,
{\rm Erf}\left(\frac{\mpl \rh}{4\,\lp P}\right)
\ ,
\ee
whose normalisation is obtained from the inner product 
\be
\pro{\psi_{\rm H}}{\phi_{\rm H}}
\equiv
\int_0^\infty
\psi_{\rm H}^*(\rh)\,\phi_{\rm H}(\rh)\,\d\rh
\ee
and can be computed numerically (for fixed $X$ and $P$).
\par
\begin{figure}[t]
\centering
\raisebox{3cm}{$|\psi_{\rm S}|^2$}
\includegraphics[width=7cm,height=3.5cm]{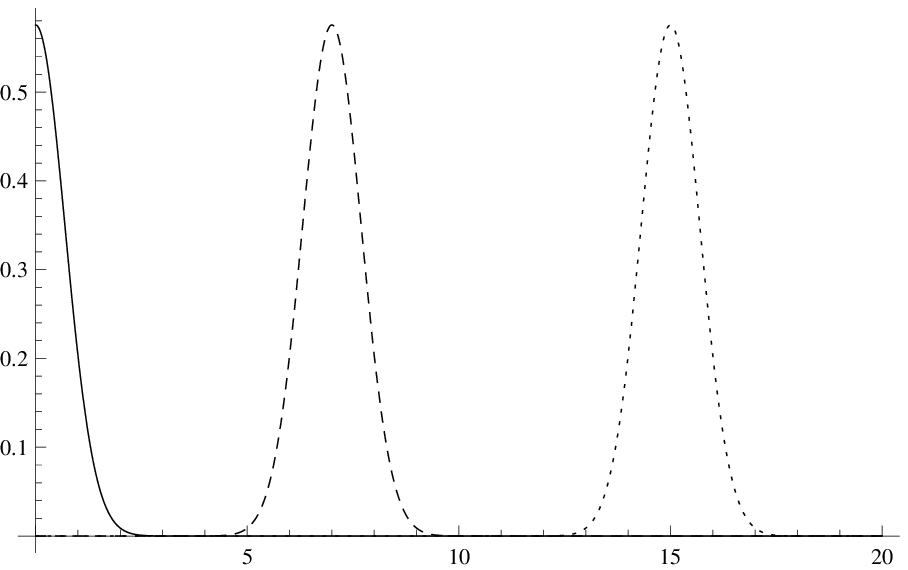}
\null
\\
\hspace{6cm}{$x/\lp$}
\\
\raisebox{3cm}{$|\psi_{\rm H}|^2$}
\includegraphics[width=7cm,height=3.5cm]{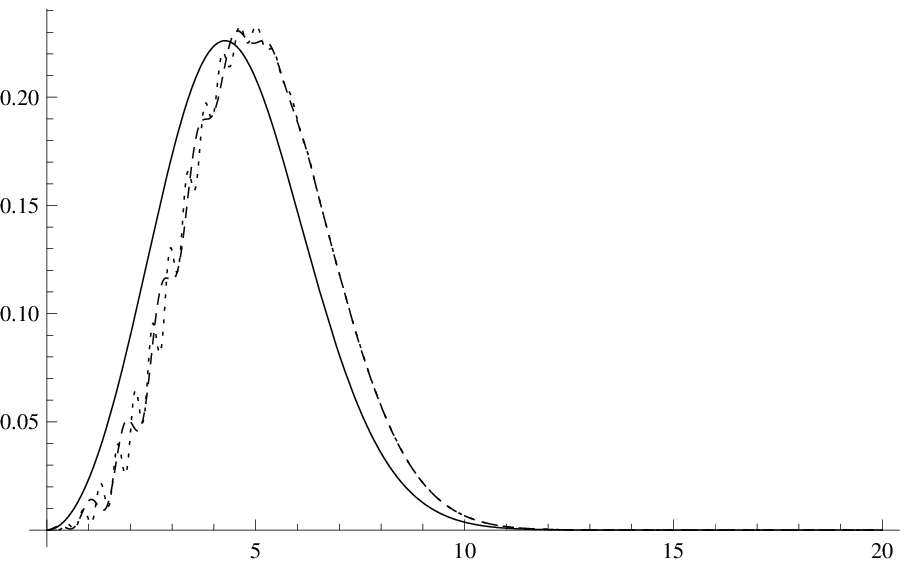}
\null
\\
\hspace{6cm}{$\rh/\lp$}
\caption{Top panel: square modulus of $\psi_{\rm S}$ for $P=\mpl$ and $X=0$ (solid line)
$X=7\,\lp$ (dashed line) and $X=15\,\lp$ (dotted line).
Bottom panel: square modulus of $\psi_{\rm H}$ for $P=\mpl$ and $X=0$ (solid line)
$X=7\,\lp$ (dashed line) and $X=15\,\lp$ (dotted line).
Particles are inside the horizon only for sufficiently small $X$.}
\label{PBHx}
\end{figure}
\begin{figure}[h!]
\centering
\raisebox{3cm}{$|\psi_{\rm S}|^2$}
\includegraphics[width=7cm,height=3.5cm]{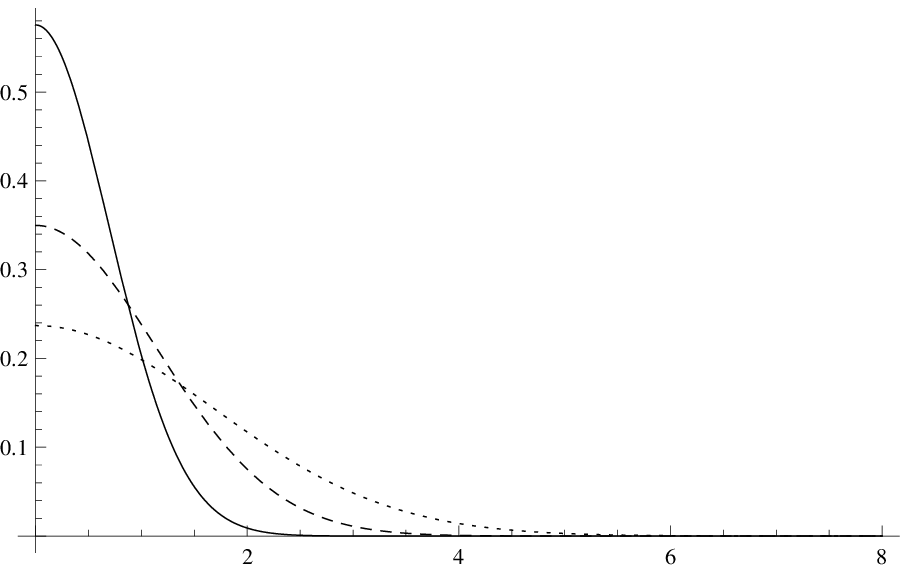}
\null
\\
\hspace{6cm}{$x/\lp$}
\\
\raisebox{3cm}{$|\psi_{\rm H}|^2$}
\includegraphics[width=7cm,height=3.5cm]{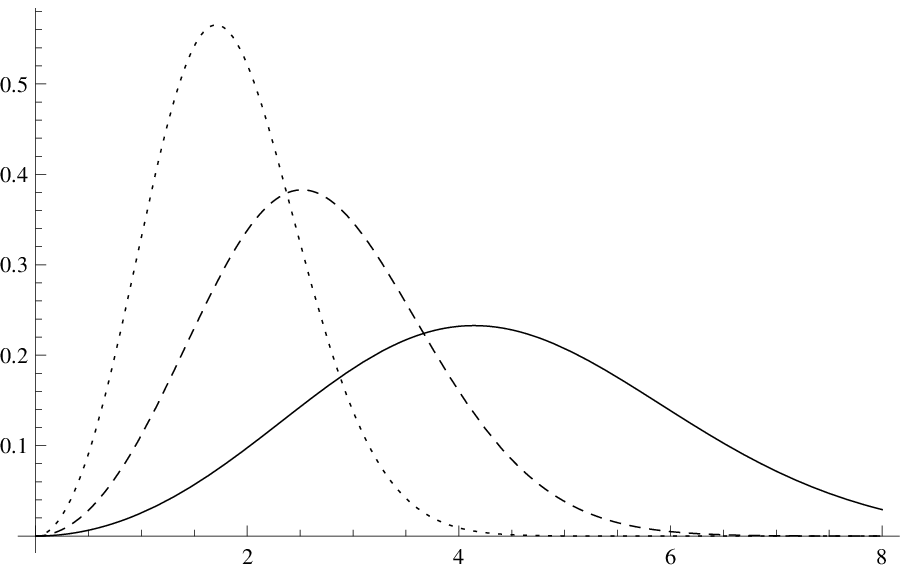}
\null
\\
\hspace{6cm}{$\rh/\lp$}
\caption{Top panel: square modulus of $\psi_{\rm S}$ for $X=0$ and $P=\mpl$ (solid line)
$P=3\,\mpl/5$ (dashed line) and $P=2\,\mpl/5$ (dotted line).
Bottom panel: square modulus of $\psi_{\rm H}$ for $X=0$ and $P=\mpl$ (solid line)
$P=3\,\mpl/5$ (dashed line) and $P=2\,\mpl/5$ (dotted line).
Particles' location is sharper the fuzzier (more spread) the horizon location
and vice versa.}
\label{PBHp}
\end{figure}
\begin{figure*}[t!]
\centering
\includegraphics[width=12cm]{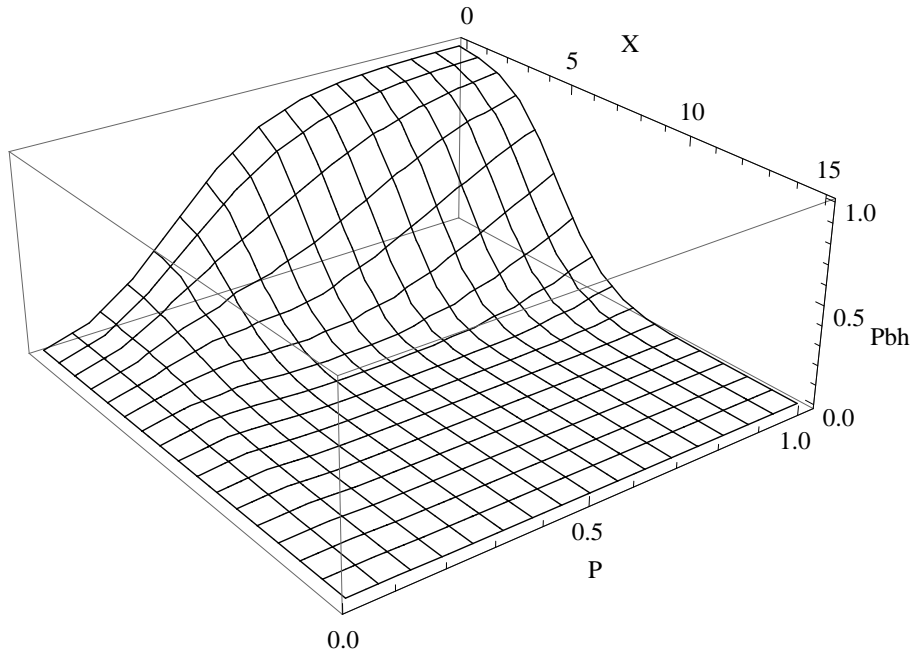}
\caption{Probability the two-particle system is a BH as a function of $X$ and
$P$ (in units of Planck length and mass respectively).}
\label{P_BH}
\end{figure*}
One then finds that $P_{\rm H}=|\psi_{\rm H}(\rh)|^2$ shows a mild dependence
on $X$ (see Fig.~\ref{PBHx}) and a strong dependence on $P$ (see Fig.~\ref{PBHp}),
in agreement with the fact that the energy of the system only depends on
$P$, and not on the spatial separation between the two particles.
It is also worth noting that $P_{\rm H}(\rh)$ always peaks around
$\rh\simeq 2\,\lp\,(2 P/\mpl)$, in very good agreement with the
hoop conjecture~\eqref{hoop}.
\par
The probability~\eqref{PBH} that the system of two particles is a BH can next
be computed numerically as a function of the distance from the centre of mass $X$
of each particle, and the total energy $2 P$.
Fig.~\ref{P_BH} shows the result for a suitable range of $X$ and $P$.
Note that a first estimate of what happens as the two particles evolve in time can be
obtained by considering the probability $P_{\rm BH}=P_{\rm BH}(X,2P)$ along lines
of constant $P$ and decreasing $X$:
$P_{\rm BH}$ clearly increases up to the maximum reached for $X=0$,
when the two (non-interacting) particles exactly superpose.
There is therefore a significant probability that the collision produces a BH,
say $P_{\rm BH}(X,2P\gtrsim 2\mpl)\gtrsim 80\%$, if the distance from the
centre of mass and linear momentum~\footnote{Recall we assumed the particles'
mass $m\ll\mpl$.}
satisfy 
\be
X
\lesssim
{2\,\lp}\,\frac{2P}{\mpl}-{\lp}
=
\rh(2P)-\lp
\ ,
\label{linearHoop}
\ee
where the term $-\lp$ on the right is the ``QM correction'' to the hoop
formula~\eqref{hoop} for $E\simeq 2P\gtrsim 2\,\mpl$, which applies to the
formation of large (semi)classical BHs.
For lower values of $P$, $P_{\rm BH}(X,2P\lesssim 2\mpl)\gtrsim 80\%$ 
for $2P\gtrsim \mpl\,(1+X^2/9\,\lp^2)$ and the limiting curve
$P_{\rm BH}(X,2P\lesssim 2\mpl)\simeq 80\%$ 
can be approximated by
\be
2P-\mpl
\simeq
\mpl\,\frac{X^2}{9\,\lp^2}
\ ,
\label{quadHoop}
\ee
which crosses the axis $X=0$ for $2P\simeq \mpl$ [instead of $2P\simeq \mpl/2$,
as it would follow from the linear relation~\eqref{linearHoop}].
Eq.~\eqref{quadHoop} represents a QM correction to the hoop conjecture~\eqref{hoop}
for quantum BH production.
Let us note that, of course, different numerical coefficients are obtained if one takes
other values of $P_{\rm BH}$ as a reasonably large probability, but the slope in
Eq.~\eqref{linearHoop} remains stably in agreement with Eq.~\eqref{hoop}.
\section{Conclusions and outlook}
Based upon the analytical results~\eqref{linearHoop} and \eqref{quadHoop},
we argue that {\em the hoop conjecture can be extended into the QM
description of BH formation\/}, with a consistent probabilistic interpretation carried
by the horizon wave-function of the given system.
It is nonetheless important to remark again that this rather neat conclusion stems
from several strong approximations.
\par
First of all, we have totally neglected any sort of interaction between the particles,
including gravitational tidal forces (which are very likely to play some part at the
Planck scale, and which will be investigated in the future).
In this respect, let us note that the flat space dispersion relation~\eqref{Ep}
might be appropriate for describing a single spherically symmetric particle,
since the relevant energy that determines its Schwarzschild radius~\eqref{hoop}
is actually the Misner-Sharp mass $M$ defined in Eq.~\eqref{M} by means of the
flat space volume measure.
Other spherically symmetric systems, such as concentric shells and spheres of
matter, are presently being analysed within this formalism.
However, once the spherical symmetry is broken, like it is for a system of two
or more colliding particles in more than one spatial dimension, the effect on
the space-time induced by each particle should be accounted for.
A way to address this issue is, for example, to employ suitably modified dispersion
relations, tantamount to a modified spectral decomposition in Eq.~\eqref{CE}.
\par
Moreover, we just considered a one-dimensional space also in order to avoid 
the kinematical complication of non-vanishing impact parameter and angular
momentum of the system of two particles.
However, it is worth recalling the original hoop conjecture should apply to
trapping surfaces, and the formation of the latter in a non-spherical massive
system has been thoroughly investigated in Ref.~\cite{malec}.
The extension of the present analysis to collisions in three spatial dimensions
is currently being investigated and we can foresee significant technical
complications.
In fact, Eq.~\eqref{hoop} and the spectral decomposition of matter states
must be replaced by the corresponding expressions for a rotating BH,
but the approach still looks promising.
\par
Let us then conclude by mentioning a further issue that has not yet been
addressed: we are all aware that, in very high energy collisions, quantum states
cannot be simply viewed as representing a fixed number of particles.
Indeed, the historical motivation for QFT was to account for particle production
and annihilation in such processes, which has made the very concept of
``localisation'' problematic. 
Since, conversely, the very concept of horizon is related to the localisation
of the matter source, one can expect that implementing the formalism of the
horizon wave-function in full-fledged QFT might lead to some surprises. 
\section*{Acknowledgments}
This work was supported in part by the European Cooperation in Science and
Technology (COST) Action MP0905 ``Black Holes in a Violent Universe".
O.M.~was supported by research grant UEFISCDI project PN-II-RU-TE-2011-3-0184.
\end{document}